\documentclass[a4paper]{easychair}
\usepackage[utf8]{inputenc}
\usepackage{xcolor}
\usepackage{listings}
\usepackage{hyperref}
\usepackage{float}
\usepackage[nomath,nosf]{kpfonts}

\definecolor{bg}{rgb}{0.95,0.95,0.95}
\newcommand{\caml}[1]{\smash{\lstinline{#1}}}
\newcommand{\cpp}[1]{\lstinline[style=C++]{#1}}

\newcommand\fnurl[2]{%
\href{#2}{#1}\footnote{\url{#2}}%
}

\usepackage{natbib}
\definecolor{britishracinggreen}{rgb}{0.0, 0.26, 0.15}
\definecolor{blue(pigment)}{rgb}{0.2, 0.2, 0.6}

\lstdefinestyle{C++}{
  language=C++,
  morekeywords={
    value,CAMLprim,CAMLdrop,CAMLreturn,
    CAMLlocal,CAMLlocal1,CAMLlocal2,CAMLlocal3,CAMLlocal4,
    CAMLparam,CAMLparam0,CAMLparam1,CAMLparam2,CAMLparam3
  },
  commentstyle=\em\color{blue(pigment)},
  stringstyle=\color{britishracinggreen},
}

\title{CAMLroot: revisiting the OCaml FFI}
\titlerunning{CAMLroot: revisiting the OCaml FFI}

\author{
  Frédéric Bour
}
\authorrunning{Frédéric Bour}

\institute{
}

\begin{document}

\maketitle

\begin{abstract}
The OCaml language comes with a facility for interfacing with C code -- the Foreign Function Interface or FFI. The primitives for working with the OCaml runtime -- and, in particular, with the garbage collector (GC) -- strive for a minimal overhead: they avoid unnecessary work and allow for calls to C code to be very cheap. But they are also hard to use properly. Satisfying the GC invariants leads to counter-intuitive C code and there are hardly any safety checks to warn the developer.

In this work, we explore two complementary approaches to mitigate these issues. First, simply adding an indirection to the API manipulating OCaml values lets us write safer code amenable to optional runtime tests that assert proper use of the API. Second, a notion of region for tracking lifetimes of OCaml values on the C side lets us trade some performance for simpler code.
\end{abstract}

\section{Introduction}

Writing code to bridge C libraries to OCaml code is a difficult task.
While writing \fnurl{Cuite}{https://github.com/let-def/cuite}, an OCaml
library that interfaces the Qt tool-kit, we discovered a few idioms that
help to keep this plumbing simple to reason about.

Qt is a C++ framework that enables writing portable user interfaces.
User interfaces are challenging to write because they involve complex
lifetimes and control flow: data is described as a dynamically changing
graph of components, control can jump back-and-forth between user code
and library code, different tasks can run concurrently, etc.

Interfacing with OCaml means exporting all these features while abiding
by OCaml \& Qt rules about memory management. By revisiting a few
assumptions of the OCaml GC interface, we believe that the CAMLroot 
approach makes interface work more tractable and easier to debug.

\subsection{Our approach: roots before values}

In our opinion, the less natural part of OCaml interaction from C code
is the manipulation of roots and the lifetime of OCaml values. The C
programmer is accustomed to manual memory management: by explicitly
creating and destroying pieces of memory or by tying the variable
lifetime to the scope. While syntactically the management of OCaml
memory seems to fall into these cases, it is actually much more subtle.

If the OCaml garbage collector triggers at the wrong time, (1) a value
can be moved, (2) a piece of OCaml memory that is locally referenced
but not registered as a root can be collected.

If (1) happens in the middle of a sequence point %
%reference the definition of C sequence point?
where the same value has been read, this results in an undefined behavior of
the C language. In practice, the lifetime of the value gets disconnected from
the lifetime of the variable that holds it.  This behavior completely
contradicts the intuition of the C developer who is not used to distinguishing
a variable from its contents.  Fortunately that can be addressed by a slight,
almost mechanical, change to the C API of the garbage collector. By making most
FFI functions take roots (in the form of pointer to values) rather than direct
values as arguments, this class of error can be ruled-out.

Furthermore, both approaches are amenable to dynamic checks that can detect even
more erroneous situations.

\subsection{Contributions}

We claim the following three contributions:
\begin{itemize}
\item a general design principle for OCaml FFI functions, working with value
      pointers rather than plain values, that prevents a class of FFI bugs and
      integrates well with existing FFI code
\item an alternative interface for managing local roots that trades some
      performance for ease of use and safety by batching roots in
      {\em regions}
\item \fnurl{camlroot}{https://github.com/let-def/camlroot} a reusable and
      open-source library that implements both of those via the \cpp{mlroot.h}
      and \cpp{mlregion.h} files.
\end{itemize}

While these were developed in the context of Cuite, the Qt binding, this
paper solely focuses on the management of OCaml memory from C code.
However to illustrate its applicability to realistic code, we describe
how our proposed API behaves in complex situations -- involving
callbacks, exceptions and multi-threaded code.

\section{The original OCaml-C FFI}

In this section we describe the existing solution for writing bindings to
external library.  The OCaml FFI lets the developer manipulate OCaml values
from another programming language. A C library provided by the OCaml
distribution exposes the primitive operations to achieve that~\citep{harmony}.

This library helps accomplish two main tasks:

\begin{enumerate}
\item constructing and deconstructing OCaml values, interpreting them in a
      meaningful way from the C language (for instance by mapping
      back-and-forth between OCaml and C representations of integers, of
      strings, etc);
\item cooperating with the OCaml garbage collector, or GC, the runtime
      service that takes care of managing OCaml memory.
\end{enumerate}

Even though both tasks are much more difficult than when working from
within OCaml code (the typechecker will not help you in foreign code),
it is at least possible to reason locally, in a compositional way, about
OCaml values -- task (1). For instance, building nested tuples just
involves repeatedly building flat tuples.
The same cannot be said about task (2). The GC needs to know about all
OCaml values that are manipulated from C code, and can look at them at
almost any moment. These restrictions are not natural while programming
in C and can lead to subtle bugs that are hard to discover.

Most of the time the GC will not do any work, preferring to wait for a
batch of work that is big enough to amortize its overhead. As such an
improper use of the OCaml API can go unnoticed for a long time. But even
once harm has been done, it might just lead to a corruption of the OCaml
heap that will affect an unrelated piece of code and fails much later in the
program.  GC bugs combine two nasty properties: they cannot be studied in
isolation and they trigger depending on a complex set of conditions that
cannot be inferred by solely looking at the buggy code.

On the other hand, the OCaml FFI API enjoys a remarkably low overhead:
the restrictions are difficult to adhere to but lead to a cheap and portable
interface with the OCaml runtime. This makes OCaml applicable to domains
where connecting to a foreign programming language is generally
considered too expensive~\citep{bourke:hal-01408230}.

We propose to explore a different trade-off in the design space of FFI
API: providing a safer and more convenient API by giving up some of the
performance.  Many mainstream languages have adopted heavier FFIs by default
(Lua, Java JNI, Go), optionally allowing to resort to a lower-level one
for performance critical code (ctypes from LuaJit); thus, a relatively
expensive FFI can still be relevant.

But before trying to build an alternative interface to the FFI, let us take a
closer look at the restrictions imposed by the GC.

\subsection{Value representation}

In C code, all OCaml values are represented by the \cpp{value} type. It is a
signed integer of the same size as a pointer of the host system (in practice,
32 or 64 bits). Values of this size are called ``words''.

The least significant bit is reserved to help the GC traverse the OCaml
heap:
\begin{itemize}
\item if it is set, the value is said to be immediate and the remaining bits
      are directly interpreted: as a 31-bit or 63-bit integer for the
      \caml{int} OCaml type, or as a unique pattern of bits for constant
      variants and polymorphic variant constructors
\item if it is not set, the value is interpreted as a pointer to a
      ``block''; it is a piece of memory provided by the runtime that is
      guaranteed to be aligned on a word boundary
\end{itemize}

Blocks are preceded by a header that determines their ``tag'' and their
size. The tag determines how to interpret the contents of the block.
For common OCaml values, such as algebraic data, records, tuples or
arrays, blocks are made of other values.

\subsection{Traversal and compaction}

Under certain conditions, the OCaml GC might need to traverse the heap.
The basic operation is to find which blocks are reachable from a value.

Depending on the tag, the OCaml GC decides whether a block is made
of other \cpp{value}s (which, in turn, can be immediate or pointer to
blocks) or just of an opaque chunk of memory that does not need to be
scanned.

By repeating this operation, the GC can traverse the whole heap. The traversal
starts from the roots, a distinguished set of values.

If necessary the GC might also decide to move some blocks.  Moving blocks is more
demanding than mere traversal: the GC not only needs to know all values
referenced from C code, it also needs to be able to update them. The C compiler
needs to be aware that all OCaml values have to be reloaded when such an
operation happens, as previous values might have been invalidated.

\subsection{The memory management macros}

A few C macros are provided by the OCaml runtime to implement foreign
features. As a guiding example, here is a simple function that takes two
values and builds a pair out of them:
\begin{lstlisting}[]
(* The OCaml version *)
let mk_pair_ocaml x y = (x,y)
(* An external function, that we will implement in C *)
external mk_pair_c : 'a -> 'b -> 'a * 'b = "mk_pair_c_impl"
\end{lstlisting}

The string after the external declaration is the name of the C function that
implements the functionality.  The corresponding C code looks like this:
\begin{lstlisting}[style=C++]
CAMLprim
value mk_pair_c_impl(value a, value b)
{
  CAMLparam2(a, b);
  CAMLlocal1(pair);
  pair = caml_alloc(2, 0);
  Store_field(pair, 0, a);
  Store_field(pair, 1, b);
  CAMLreturn(pair);
}
\end{lstlisting}

The first macro \cpp{CAMLprim} ensures that the symbol is visible from OCaml
code.

\paragraph{CAMLparam}
The \cpp{CAMLparam2(a,b)} call expands to two other macros:

\begin{itemize}
  \item \cpp{CAMLparam0()} saves the previous set of local roots 
  \item \cpp{CAMLxparam2(a,b)} setups a new block of roots with the
      addresses of \cpp{a} and \cpp{b}.
\end{itemize}

The local roots are in a linked list of pointers to OCaml values,
implemented by the \cpp{struct} \cpp{caml\_\_roots\_block} type, and
stored in the \cpp{caml\_local\_roots} global variable.

The job of the memory management macros is to make it as easy as
possible to register all local variables of type \cpp{value} in this
linked list and to remove them when returning from the function.

There should be only one \cpp{CAMLparam0()} in a function, but there
can be as many calls to \cpp{CAMLxparam} as needed. The variants from
\cpp{CAML(x)param1} to \cpp{CAML(x)param5} are available as well as
\cpp{CAML(x)paramN(array,\ array\_size)} for registering array of values.

\paragraph{CAMLlocal} The next macro call of interest is
\cpp{CAMLlocal1(pair)}. 

It expands to \cpp{value\ pair\ =\ Val\_unit;\ CAMLxparam1(pair)}:
\begin{itemize}
\item it declares and initializes a local variable named \cpp{pair},
\item it adds its address to the set of local roots.
\end{itemize}

The next lines, the calls to \cpp{caml\_alloc} and
\cpp{Store\_field}, are not directly related to the management of
roots. They deal with the construction of OCaml values -- assuming that
all variables have been registered properly.

\begin{figure}[htbp]
\begin{lstlisting}[style=C++]
// Allocating values
value caml_alloc(int size, int tag);
value caml_copy_string(const char *string);
...
    
// Deconstructing and mutating values (actually implemented by macros)
long Long_val(value v);
value Val_long(long x);
value Field(value v, int offset); 
void Store_field(value v, int offset, value x);
...
\end{lstlisting}
  \caption{Some OCaml FFI functions for manipulating values}
  \label{fig:mlvalue}
\end{figure}

\paragraph{CAMLreturn} This last macro restores the previous set of local
roots. It sets the variable \cpp{caml\_local\_roots} to the state that was
saved by \cpp{CAMLparam0()}.

Thus, the code above can be desugared to the equivalent:
\begin{lstlisting}[]
CAMLprim
value mk_pair_c_impl(value a, value b)
{
  // CAMLparam2(a, b);
  CAMLparam0();        // 1) save the state of local roots
  CAMLxparam2(a, b);   // 2a) add &a and &b to local roots

  // CAMLlocal1(pair);
  value pair = Val_unit;
  CAMLxparam1(pair);   // 2b) add &pair to local roots

  ...
  
  // CAMLreturn(pair);
  CAMLdrop;            // 3) restore the state of local roots
                       //    (forgetting &a, &b and &pair)
  return(pair);
}
\end{lstlisting}

The three fundamental operations of root managements are saving local roots,
registering new ones, and restoring the saved ones when leaving a scope.

The OCaml FFI provides macros to automate most of this work but has no way to enforce their proper use. \cpp{mlroot.h} API can detect large classes of possible misuses while \cpp{mlregion.h} introduces an alternative approach to the management of roots. 

\subsubsection{Carefully dealing with intermediate results}

Here is an example that shows how easy it is to misuse this API, taken from
\citep[caml-oxide]{camloxide}.  Lets imagine one wants to make a triplet as two nested pairs:
\begin{lstlisting}[]
let triplet x y z = (x,(y,z))
\end{lstlisting}

Armed with \cpp{mk\_pair\_c\_impl} and the rules above, one might 
be tempted to write:
\begin{lstlisting}[style=C++]
CAMLprim
value c_triplet(value x, value y, value z)
{
  CAMLparam3(x,y,z);
  CAMLlocal1(triplet);

  triplet = mk_pair_c_impl(x, mk_pair_c_impl(y, z));

  CAMLreturn(triplet);
}
\end{lstlisting}

But a bug lies in this implementation: the C compiler might have already
loaded the value of \cpp{x} (for instance, by copying it on the
stack) before the nested call to \cpp{mk\_pair\_c\_impl(y,\ z)}.

If this call triggers a compaction and \cpp{x} is moved, the old,
and wrong, value of \cpp{x} will be passed to the outer call.

The correct version uses an intermediate variable for the temporary
value:
\begin{lstlisting}[style=C++]
CAMLprim
value c_triplet(value x, value y, value z)
{
  CAMLparam3(x,y,z);
  CAMLlocal2(intermediate, triplet);

  intermediate = mk_pair_c_impl(y, z);
  triplet = mk_pair_c_impl(x, intermediate);

  CAMLreturn(triplet);
}
\end{lstlisting}

To avoid bugs, calls to functions manipulating the OCaml memory should be
linearized and temporary results should be stored in local roots.

\section{mlroot: solving problems with one level of indirection}

The first change we propose is to replace the type of arguments of type \cpp{value}
to the type \cpp{value*}, representing roots rather than direct values.
Similarly, return values of type \cpp{value} are replaced by an extra argument
of type \cpp{value*}, which will be used to store the results.

This rewriting is only necessary for functions that allocate, but for the sake
of uniformity we offer alternatives in this style for most GC functions.
Figure~\ref{fig:mlvalue} shows some functions from the original OCaml FFI while
figure~\ref{fig:mlroot} shows the equivalent functions provided by mlroot API.
This minor change brings many benefits.

\paragraph{No risk of unexpected copy.}

A tricky source of bug that we highlighted in the previous section was
that OCaml values can be unexpectedly copied while being manipulated. With
the added indirection only the pointer is copied. If the GC kicks in and
rewrites the roots, the pointer will not be affected anyway.

Looking at the operations involved in terms of lifetime, reading a value
from a root makes it ephemeral: the value is valid only until the next
OCaml allocation -- or simply undefined if an allocation can happen in
the same sequence-point.

When directly working with \cpp{value}, this operation is implicit.
Working with \cpp{value*}, the operation becomes explicit and forces
the developer to think about its effects -- they choose when to
dereference the pointer.
In practice this is almost never needed outside the implementation of
\cpp{mlroot}, relieving the user of the API from this burden.

\begin{figure}[htbp]
\begin{lstlisting}[style=C++]
// Allocating values
void mlroot_alloc(value *root, mlsize_t size, tag_t tag);
void mlroot_string_copy(value *root, const char *string);
...
    
// Deconstructing and mutating values
long mlroot_get_long(value *root);
void mlroot_set_long(value *root, long x);
void mlroot_get_field(value *root, const value *src, int index);
void mlroot_set_field(const value *root, int index, const value *src);
...
\end{lstlisting}
  \caption{Functions from \ref{fig:mlvalue} following mlroot conventions}
  \label{fig:mlroot}
\end{figure}

\paragraph{Preventing nested calls.}

Now that all functions that interact with the garbage collector take pointers
and return void, offending code patterns become much harder. Calls cannot be
nested anymore.

Here is what the \cpp{triplet} would look like with this approach:
\begin{lstlisting}[style=C++]
static void mk_pair(value *result, value *a, value *b)
{
  mlroot_assert(result != a && result != b);
  mlroot_alloc(result, 2, 0);
  mlroot_set_field(result, 0, a);
  mlroot_set_field(result, 1, b);
}

CAMLprim
value caml_triplet(value x, value y, value z)
{
  CAMLparam3(x, y, z);
  CAMLlocal2(pair, result);
  mk_pair(&pair, &y, &z);
  mk_pair(&result, &x, &pair);
  CAMLreturn(result);
}
\end{lstlisting}

\paragraph{No need to repeat roots.}

Callees no longer have the responsibility of registering roots for their arguments.

With the existing OCaml API, any function receiving an argument of type
\cpp{value} has to register a corresponding root. There are as many
roots for the same value as sub-routines calls that received it as
argument.  With the indirect approach only places that have their
address taken need to be registered as root. Since this operation is explicit, we believe the risk of mistake is reduced.

Dereferencing should also be done with care, but this will generally
be done by a primitive function of the GC interface and not by the user code anymore.

\paragraph{Dealing with immediate values.}

A reader familiar with OCaml binding code might be worried that working
with immediate values (an integer directly stored in a \cpp{value})
becomes less convenient with our approach than with the normal API.

Immediate values enjoy a lot of nice properties in the OCaml FFI. Since
they do not interact with the memory graph of OCaml -- they don't reference
blocks, they cannot be moved -- the rules for dealing with them are relaxed:
they don't have to be put in roots, they can be created without triggering a
garbage collection, etc.

We argue these properties should not be exploited. That some values can be
represented without interacting with the GC is an implementation detail. Being
prepared for the ``worst case'' allow to present a uniform interface.

Still, the library provides the \cpp{mlroot\_val\_long} and
\cpp{mlroot\_set\_field\_long} short-hands for cases that are known to be
safe.

\subsection{Safety of this indirect API}

Moving everything to pointers opens a new opportunity for incorrect
uses: aliasing. In the \cpp{triplet} case it would mean using the
\cpp{result} variable both as input and output in the same call:
\begin{lstlisting}[style=C++]
CAMLprim
value caml_triplet(value x, value y, value z)
{
  CAMLparam3(x, y, z);
  CAMLlocal1(result);
  mk_pair(&result, &y, &z);
  mk_pair(&result, &x, &result); // result is aliased!
  CAMLreturn(result);
}
\end{lstlisting}

While problematic indeed, this case is actually less worrying. The code
that dereferences roots can be instrumented to deal with that:
\begin{itemize}
\item by properly handling aliasing, for instance by ensuring that all
      arguments are read before any are written,
\item by checking for this case and failing or emitting a warning, as
      illustrated by the assertions in the implementation of
      \cpp{mk\_pair} primitive. 
\end{itemize}

\paragraph{A debugging workflow.} Actually thanks to the indirection we can go
further than that. The observation is that any well-formed argument of type
\cpp{value*} should point to a root. 

The native OCaml FFI is \texttt{value}-centric: functions directly take and produce
values. Our rewriting make it \texttt{root}-centric: functions receive roots
and manipulate values though them.

With native OCaml FFI the connection between a root and its value is lost: when
a value argument is passed, a copy is made and it is not possible to know which
root it got copied from.  However \cpp{mlroot} functions check that these roots
have been properly registered by plugging into the GC infrastructure. This
comes at a moderate speed cost in the form of a "defensive" mode that can be
switched on during development.

\paragraph{Where to add the indirection?}

Having made explicit the distinction between values (of type
\cpp{value}) and roots (of type \cpp{value*}) in the API, one
could wonder why our API makes use of roots in places where values would
be fine: the arguments to \cpp{mlroot\_get\_field}, \cpp{mlroot\_long\_val},
etc.  We are not totally decided on this issue and might revisit this design
in the future. However the ability to dynamically check for correct use
and the more explicit, safer-looking nature of the resulting code makes
us favor the root arguments.

Beside the slight increase in verbosity, we did not find any drawback to
this approach.  The indirection does not increase memory use because the root
has to be registered in the GC anyway. The impact on execution time is not significant either: the only change is that the pointer dereferencement (a memory load) is done eagerly with the value-centric API while it is delayed with the root-centric approach.

\section{mlregion: dynamic allocation of roots}

To further simplify the API described above we propose to make
allocation of roots simpler.

The \cpp{CAMLparam} and \cpp{CAMLlocal} macros declare OCaml roots
with a static lifetime, known at compile-time. This is nice for
performance but puts more burden on the developer.

The semantics of these macros is hard to understand and some use cases
are not easily covered. As we already saw, returning values is tricky,
but storing temporary values in code controlled by an external framework
is even more problematic.

\paragraph{The need for side-channel allocation.}

For the sake of the example, let's imagine that we need to sort some C
structures containing OCaml values. To achieve this the
\cpp{qsort\_r} function from the C standard library seems appropriate. It
takes an array of user-defined structures and a custom comparison
operator in the form of a function pointer.
\begin{lstlisting}[style=C++]
struct item {
  my_c_type x;
  value v;
};

static int
c_comparator(const void *item1, const void *item2, void *comparator)
{
    value v1 = ((const struct item *)item1)->v;
    value v2 = ((const struct item *)item2)->v;
    value ml_comparator = *(value*)comparator;
    return Val_int(caml_callback2(ml_comparator,v1,v2));
}

void sort_ocaml_items(struct item *items, size_t count, value *comparator)
{
    qsort_r(items, count, sizeof(struct item),
            c_comparator, comparator);
}
\end{lstlisting}

\cpp{caml\_callback2} is a primitive function of native OCaml FFI
that allows to invoke an OCaml closure from C code\footnote{For the sake of
simplicity we do not deal with the case where the callback raises an
exception}.

Because of this callback, the garbage collector can be called in the
middle of the sorting. Even if we registered roots for all the values
in this array, the implementation of \cpp{qsort\_r} might have made
copies that will not be updated by the GC. More generally, rewriting
the array in the middle of the sort can lead to unexpected behaviors.

Since we know all the OCaml values that will be reached prior to calling
\cpp{qsort\_r}, a solution is to work with pointer to values. One
first allocates an array of roots and passes pointers into this array.

However there exist situations where the set of roots cannot be pre-determined.
Regions appeared as a solution to this problem, and proved to be convenient in
simpler cases too.

\subsection{Region-based management}

To let the developer dynamically manage the set of roots, we propose a
simple API that over-approximates the lifetime of local roots:
\begin{lstlisting}[style=C++]
typedef struct ... region_t;
void mlregion_enter(region_t *region);
void mlregion_leave(region_t *region);
value *mlregion_new_root(void);
#define CAMLregion(...) ...
#define CAMLregion_return(p) ...
\end{lstlisting}

In this approach, we distinguish between external and helper functions:
\begin{itemize}
\item external functions are the ones that can be directly called from OCaml,
\item helper functions implement useful routines for binding foreign code.
\end{itemize}

The external functions are responsible for setting up the region while
helper functions assume that a region has already been set up. Mimicking
\cpp{CAMLparam...} macros, we provide some helpers for registering
parameters while setting up the region:
\begin{lstlisting}[style=C++]
value *pair_helper(value *a, value *b)
{
    value *v = mlregion_new_root();
    mlroot_alloc(v, 2, 0);
    mlroot_set_field(v, 0, a);
    mlroot_set_field(v, 0, b);
    return v;
}

CAMLprim
value mk_pair(value a, value b)
{
    CAMLregion(&a, &b);
    CAMLregion_return(pair_helper(&a, &b));
}

CAMLprim
value mk_triplet(value x, value y, value z)
{
    CAMLregion(&x, &y, &z);
    CAMLregion_return(pair_helper(&x, pair_helper(&y, &z)));
}
\end{lstlisting}

Setting up a region introduces a new set of local roots that can grow
dynamically as new roots are requested. Leaving a region releases all
the roots at once.

\paragraph{Dynamic scoping of regions.} A point that might surprise users of
this API is that the current region is not explicitly passed to functions,
instead it is accessed by some external means.

This design choice was made to simplify integration with Qt code: OCaml code
can get called from a method deep in the object hierarchy, whose interface is
imposed by the framework. As there is no easy way to thread a region handle to
that point, dynamic scoping comes naturally as a solution. We might revisit this
decision later. For instance regions could be threaded explicitly by default,
and auxiliary functions could allow to set and retrieve the current region for
situations where threading is not possible.

\subsection{Sub-regions}

Assuming that all roots have the same life-time as the external
entrypoint works well if a fixed amount of work has to be done. However,
for long-running function (for instance, an event loop driven by
C-code), the over-approximation of lifetimes can be problematic. For
these cases, we allow the introduction of sub-regions, valid in a local
scope.

These sub-regions follow a stack discipline: they can be nested and are
released in the reverse order of their allocation.
\begin{lstlisting}[style=C++]
void mlregion_subenter(region_t *region);
void mlregion_subleave(region_t *region);
\end{lstlisting}

For instance, the following code avoids leaking roots while transforming
all the elements of an array:
\begin{lstlisting}[style=C++]
void process_item(value *acc, value*item);

void fold_array(value *acc, value *array)
{
    region_t region;
    size_t count = mlroot_get_size(array);
    for (size_t i = 0; i < count; ++i)
    {
        mlregion_subenter(&region);
        value *item = mlregion_new_root();
        *item = mlroot_get_field(array, i);
        process_item(*acc, *item);
        mlregion_subleave(&region);
    }
}
\end{lstlisting}

Macros can be used to automate some of the boilerplate.

\subsection{Releasing the lock in a region}

So far we have demonstrated the use of regions to allocate and manage OCaml
memory. The concept can also be applied to the converse: preventing
allocation and manipulation of OCaml memory in a given scope.

Although a multi-core runtime is being developed~\citep{multicore}, the vanilla
OCaml runtime can only execute on a single thread of execution. When multiple C
threads are in use, a lock is used by the OCaml runtime to ensure that only one
of them executes OCaml code at any given time.

The C FFI provides an API for releasing the OCaml runtime lock in a given
scope of code.
\begin{lstlisting}[style=C++]
// Existing API
void caml_release_runtime_system(void);
void caml_acquire_runtime_system(void);
\end{lstlisting}

These APIs can be wrapped in corresponding
\cpp{mlregion\_\{acquire,release\}\_runtime\_system} functions that does 
additional bookkeeping to ensure proper use of regions while the runtime is
released:
\begin{itemize}
\item new roots cannot be allocated,
\item dereferencing a value is forbidden, most helper functions won't work,
\item setting up normal regions is forbidden, but a special kind of region
      allows reacquiring the runtime.
\end{itemize}

All these restrictions can be tested at a moderate cost. While no
checks are done at compile-time, misuse of the API can be reliably detected
during execution.
\begin{lstlisting}[style=C++]
// Wrappers for releasing the runtime
void mlregion_release_runtime_system(void);
void mlregion_acquire_runtime_system(void);

// Wrappers for locally reacquiring the runtime
void mlregion_reacquire_runtime_system(void);
void mlregion_rerelease_runtime_system(void);
\end{lstlisting}

\subsection{Calling OCaml from region-managed code}

The last feature that needs some special care from the region API is the
ability to call OCaml closures from C code. When switching back to OCaml
code, the runtime marks a region as disabled: the roots it contains are
still reachable, but no new roots can be added to the region.

This helps detect and handle a few unfortunate cases:
\begin{itemize}
\item When re-entering C code from OCaml deeper in the call stack, an
      entrypoint that forgot to setup a region could allocate from the outer
      region by mistake.
\item If we are unlucky, the OCaml thread scheduler could preempt the
      current thread and the re-entry would happen from another
      thread, damaging the internal datastructures of the regions library. By
      wrapping calls with custom code, we can rely on the OCaml runtime lock to
      also protect region sections.
\item The OCaml closure could raise an exception. The native FFI deals with
      this situation by simply dropping roots from the local roots linked
      list: since the nodes allocated by \cpp{CAMLparam/local} macros are
      stored on the stack, when an exception is raised the local root and
      stack pointers are simply reset to their value before entering the C
      code. A workaround for regions is discussed below.
\end{itemize}

\paragraph{Handling exceptions.}

The OCaml native FFI provides two means for calling OCaml closures:
\begin{itemize}
\item the \cpp{CAMLcallback()} variants, that do not intercept
      exceptions. The C code will be aborted by directly jumping to the
      OCaml code that called an external function.
\item the \cpp{CAMLcallback\_exn()} variants, that tag the return value
      to distinguish exceptional case.
\end{itemize}

The return value of \cpp{CAMLcallback\_exn()} should be tested for
the exceptional case with \cpp{Is\_exception\_result} before resuming
normal execution.  Because our region management code needs to execute cleanup
code when leaving a scope, we forbid the former case. The user-code has to
handle the exceptional case without resorting to non-local control flow.

While it would have been possible to provide support for non-local
jumps, it did not made much sense in the Qt case: the binding is
implemented in C++, which allow arbitrary code to be executed when
leaving a scope. C++ exceptions are expressive enough to handle all our
requirements (non-local control flow, proper interaction with the
regions and with OCaml GC), but the bindings themselves did not need
that feature.

\section{Future work}

\cpp{mlroot} and \cpp{mlregion} emerged during the design of the Cuite library
and are extracted from its core code.  As the project is still in its infancy,
it is evolving rapidly and the libraries have been properly tested only for the
use cases stressed by Cuite. We still have to cover the rest of the FFI API.

Similarly, the support for runtime checks was only used for a few ad-hoc
cases. Devising and implementing a robust suite of dynamic checks that are
useful beyond Cuite is on our roadmap. Thanks to the transparent integration
with the original FFI, this would help to debug existing bindings.

As Cuite is developped in C++, we have already developped a C++ layer on top of
the C API to make it more idiomatic and remove some of the boilerplate --
using references instead of pointers for dealing with roots, using
RAII-idiom~\citep{Stroustrup:1995:DEC:193198} to ensure well-bracketed nesting
of regions, etc.
Extracting and generalizing this part would extend the usefulness of our
library to other C++ bindings.

Finally, the only part of the OCaml runtime that we rely on and that is not
already in the public interface is the representation of local roots. 
This part has already been stable for years. We are quite confident that it
will be possible to get guarantees that this API will not break in future
releases of OCaml.

\section{Related Work}

The safety and simplicity of foreign function interfaces for OCaml has been
approached from many angles.

O-Saffire~\citep{Furr:2005:CTS:1064978.1065019} is a static analysis that works
on the official OCaml FFI. It goes beyond checking the registration of roots
and also checks that the value representation on the C and OCaml sides is
compatible.

Unfortunately, O-Saffire has not received much changes since 2005 and we could
not get it to work on a recent distribution of OCaml.

Ctypes~\citep{DBLP:journals/scp/YallopSM18} proposes an alternative way to bind
libraries. Rather than writing C code, a specification of the library is
described in OCaml code. From this specification bridging code will be
generated. The code can be instrumented to check for different safety
properties.

Ctypes is very convenient for binding simple C functions. Since most of the
code is described using OCaml combinators, it is easy to get bindings that are
type-safe by construction.  However it falls short on two fronts:  there is
limited support for calling C++ code and for manipulating objects with complex
lifetimes or custom memory management rules.  In these cases one has to write
low-level code that follows the requirement of the library being bound.
Achieving that through Ctypes combinators can prove more cumbersome and limited
than directly writing the corresponding C code.

Caml-oxide~\citep{camloxide} is a proof of concept implementation of an OCaml
FFI for Rust. In particular, it demonstrates that the restrictions applying to
GC roots can be tracked by Rust type system.

This is the most promising approach for getting bindings that are safe by
construction. It does so by leveraging the type systems of OCaml and of Rust.
As such, it cannot help with C/C++ libraries. The actual implementation is
also too limited for most practical applications: it only covers a minimal part
of the GC API, just enough to demonstrate the viability of the approach.

% \section{Conclusion}
% 
% Focusing on roots rather than values 
% Revisiting a few assumptions of the OCaml foreign function interface can help
% solving some difficult problems.
% 
% materializes roots as a different type, here \cpp{value*}
%%%%%%%%%%%%%%%%%%%%%%%%%%%%%%%%%%%%%%%%%%%%%%%%%%%%%%%%%%%%%%%%%%%%%%%%%%%%%%

\label{sect:bib}
\bibliography{CAMLroot}

\end{document}